\documentclass[
superscriptaddress,
showpacs,preprintnumbers,
 amsmath,amssymb,
 aps,
prapplied,
]{revtex4-2}

\usepackage{graphicx}
\usepackage{dcolumn}
\usepackage{bm}
\usepackage[update,prepend]{epstopdf}
\usepackage{xcolor}
\usepackage{ulem}
\usepackage{hyperref}

\begin{document}

\preprint{Arxiv version}

\title{Strong coupling of exciton-polaritons in a bulk GaN planar waveguide: quantifiying the { coupling strength } }

\author{C.~Brimont}
\affiliation{Laboratoire Charles Coulomb (L2C), Universit\'e de Montpellier, CNRS, Montpellier, France}
\author{L.~Doyennette}
\affiliation{Laboratoire Charles Coulomb (L2C), Universit\'e de Montpellier, CNRS, Montpellier, France}

\author{G.~Kreyder}
\affiliation{Universit\'e Clermont Auvergne, CNRS, SIGMA Clermont, Institut Pascal, Clermont-Ferrand, France}
\author{F.~R\'everet}
\affiliation{Universit\'e Clermont Auvergne, CNRS, SIGMA Clermont, Institut Pascal, Clermont-Ferrand, France}
\author{P.~Disseix}
\affiliation{Universit\'e Clermont Auvergne, CNRS, SIGMA Clermont, Institut Pascal, Clermont-Ferrand, France}
\author{F.~M\'edard}
\affiliation{Universit\'e Clermont Auvergne, CNRS, SIGMA Clermont, Institut Pascal, Clermont-Ferrand, France}
\author{J.~Leymarie}
\affiliation{Universit\'e Clermont Auvergne, CNRS, SIGMA Clermont, Institut Pascal, Clermont-Ferrand, France}

\author{E.~Cambril}
\affiliation{Centre de Nanosciences et de Nanotechnologies, CNRS, Universit\'e Paris-Saclay, France}
\author{S.~Bouchoule}
\affiliation{Centre de Nanosciences et de Nanotechnologies, CNRS, Universit\'e Paris-Saclay, France}

\author{M.~Gromovyi}
\affiliation{Centre de Nanosciences et de Nanotechnologies, CNRS, Universit\'e Paris-Saclay, France}
\affiliation{UCA, CRHEA-CNRS, Rue Bernard Gregory, 06560 Valbonne, France}
\author{B. Alloing}
\affiliation{UCA, CRHEA-CNRS, Rue Bernard Gregory, 06560 Valbonne, France}
\author{S.~Rennesson}
\affiliation{UCA, CRHEA-CNRS, Rue Bernard Gregory, 06560 Valbonne, France}
\author{F.~Semond}
\affiliation{UCA, CRHEA-CNRS, Rue Bernard Gregory, 06560 Valbonne, France}
\author{J.~Z\'u\~niga-P\'erez}
\affiliation{UCA, CRHEA-CNRS, Rue Bernard Gregory, 06560 Valbonne, France}

\author{T.~Guillet}
\email{Thierry.Guillet@umontpellier.fr}
\affiliation{Laboratoire Charles Coulomb (L2C), Universit\'e de Montpellier, CNRS, Montpellier, France}
\affiliation{Department of Physics, SUPA, University of Strathclyde, Glasgow, United Kingdom}

\date{\today}

\begin{abstract}

We investigate the demonstration and quantification of the strong coupling between excitons and guided photons in a GaN slab waveguide. The dispersions of waveguide polaritons are measured from T=6~K to 300~K through gratings. They are carefully analyzed within four models based on different assumptions, in order to assess the strong coupling regime. We prove that the guided photons and excitons are strongly coupled at all investigated temperatures, with a small $(11 \%)$ dependence on the temperature. However the values of the Rabi splitting strongly vary among the four models: the ``coupled oscillator'' model over-estimates the coupling; the analytical ``Elliott-Tanguy'' model precisely describes the dielectric susceptibility of GaN near the excitonic transition, leading to a Rabi splitting equal to $82 \pm 10 \ meV$ for TE0 modes; the experimental ellipsometry-based model leads to smaller values of $55 \pm 6 \ meV.$ 
{ We evidence that for waveguides including active layers with large oscillator strengths, as required for room-temperature polaritonic devices, a strong bending of the modes dispersion is not necessarily the signature of the strong-coupling, which requires for its reliable assessment a precise analysis of the material dielectric susceptibility.}

\end{abstract}

\maketitle


\section{\label{sec:intro}Introduction}

Exciton-polaritons are the dressed states of semiconductor excitons when they interact with photons in the strong coupling  regime. Compared to photons, the interactions between polaritons are enhanced by a few orders of magnitude \cite{baas_optical_2004, vladimirova_polariton-polariton_2010}, allowing for the exploration of a wide range of collective quantum states and phenomena \cite{kasprzak_bose-einstein_2006, lagoudakis_observation_2009, lagoudakis_coherent_2010, amo_polariton_2011} inspired from the physics of cold atoms. Most of these studies have been performed on microcavity-polaritons, where 2D photons are confined in a vertical Fabry-Perot cavity \cite{kavokin_microcavities_2006}. More complex polariton states can be engineered through photonic circuits \cite{Amo_Exciton_2016, wertz_spontaneous_2010, gao_polariton_2012} or optical control of the excitation pattern \cite{lagoudakis_coherent_2010, cristofolini_optical_2013}. Alternative geometries have been recently explored, where excitons are coupled to Tamm plasmons \cite{symonds_emission_2009,kaliteevski_hybrid_2009}, Bloch surface states \cite{liscidini_guided_2011}, micro- and nanowires \cite{van_vugt_exciton_2006, czekalla_whispering_2008, zimmler_optically_2010, zhang_robust_2012} and guided modes in a slab \cite{walker_exciton_2013}.

The waveguide geometry provides an interesting framework for polariton fluids propagating with a large group velocity, and with a strong nonlinearity, leading to the formation of bright or dark solitons carrying a very low energy per pulse of about one $pJ$ { in GaAs-based waveguides at cryogenic temperatures\cite{walker_ultra-low-power_2015, walker_dark_2017}, and pulse amplitude modulation $\sim 100 pJ$ in GaN-based waveguides at room temperature\cite{Paola_Ultrafast_2020}.} 
The waveguide geometry was recently investigated in GaN and ZnO semiconductors in order to achieve large Rabi splittings \cite{oder_propagation_2001, solnyshkov_optical_2014, ciers_propagating_2017, jamadi_edge-emitting_2018}, attaining up to about $200 \ meV$ for a ZnO slab waveguide and enabling the demonstration of a polariton laser and a polariton amplifier operating up to room temperature \cite{ jamadi_edge-emitting_2018}.
Indeed, room temperature polaritonics requires robust polaritons involving  excitons with a large oscillator strength and a large binding energy, found either in wide bandgap semiconductors or in organic materials \cite{guillet_polariton_2016}. When large Rabi splittings are achieved in these systems, a proper measurement and modelling of the dispersion of the eigenmodes is necessary, a much more difficult task than when dealing with standard semiconductors (e.g. GaAs) and  a vertical geometry \cite{richter_maxwell_2015, ciers_propagating_2017}. { Indeed, on the experimental side, the damping of the upper polariton branch by the absorption of continuum states frequently precludes its observation and the direct measurement of the Rabi splitting between lower and upper polariton states \cite{faure_comparison_2008, reveret_modelling_2010}; on the theoretical side,} refractive index contributions from bound excitons, unbound scattering states and continuum states are a long-debated issue \cite{elliott_intensity_1957, tanguy_optical_1995}, which becomes prominent when investigating wide bandgap semiconductors and organic materials \cite{djurisic_progress_2002}. This is due to the large corrections they introduce to the so-called background refractive index. The role of these contributions was already evidenced in the dispersion of polaritons in vertical microcavities based on wide bandgap materials \cite{guillet_laser_2011,guillet_polariton_2016}.

{ In general, the precise determination of the exciton-photon coupling strength requires the identification of the condition of zero-detuning, where the  energies of the bare photon mode (without coupling to the excitons) and of the exciton are equal. Extracting such a simple condition from the experimental dispersions becomes a challenging task due to the difficulty to answer the question: what would be the dispersion of a bare waveguide mode in a gedanken experiment where exciton oscillators would be ``turned off'' in the optical response of the active layer? 
To answer this question, we measure the polariton dispersions in a bulk GaN polariton waveguide from low to room-temperature and propose four models to simulate them, extracting therefrom the model-dependent exciton-photon coupling strengths. These models go from the most basic and most-frequently employed coupled-oscillators model to more elaborate ones that implement more accurate descriptions of the materials dielectric susceptibility.
We compare and analyze critically the very different estimates of the exciton-photon coupling strength while fitting the same experimental polariton dispersions.}

\section{\label{sec:sample}Sample}

\begin{figure}
\resizebox{\hsize}{!}{\includegraphics{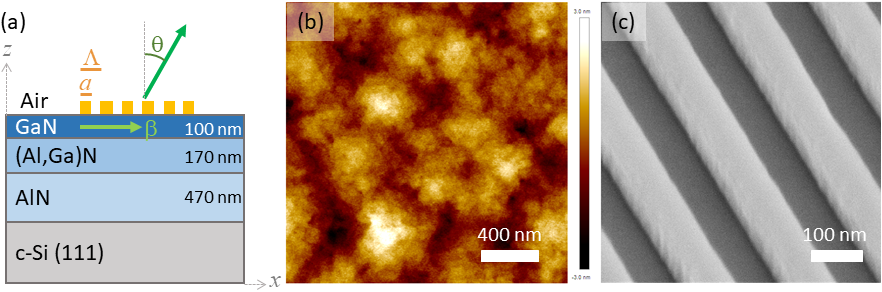}} 
  \caption{ (a) Sample structure; $\beta$ represents the propagation constant of the guided polaritons, and $\theta$ is the outcoupling angle; (b) Atomic Force Microscopy image (AFM) of the top GaN surface (z-scale $6 \ nm;$ RMS roughness $0.8 \ nm);$ (c) Scanning electron micrograph (SEM) of the etched diffraction grating.}
  \label{fig:sample}
\end{figure}

The structure schematically depicted in Fig.~\ref{fig:sample}.a has been grown by molecular beam epitaxy on a Si~(111) substrate. The epitaxy of a thick $(470 \ nm)$ AlN buffer at high temperature $(1000 ^o C)$ is followed by a $170 \ nm-$thick Al$_{0.65}$Ga$_{0.35}$N, before depositing the GaN waveguide core. The AlN and  Al$_{0.65}$Ga$_{0.35}$N cladding layers prevent the overlap between the waveguide photonic mode and the Si substrate, which would otherwise result in a strong absorption. The topmost surface of the GaN waveguide displays a surface roughness as low as $1 \ nm \ RMS$, as illustrated in Fig.~\ref{fig:sample}.b.
The typical thickness of the core GaN layer{, measured in scanning electron micrographs (SEM),} is $100 \ nm$.
The sample is almost crack-free, with a typical distance between cracks of the order of several millimeters.

First-order grating couplers have been defined by electron beam lithography (Fig.~\ref{fig:sample}.c) using a negative-tone hydrogen silsesquioxane resist  (HSQ~-~flowable oxide). After HSQ resist development, the sample is baked under O$_2$ atmosphere at $440 ^o C$ for 1~hour to stabilize the HSQ refractive index to a value of $n \approx 1.43$ and reduce the extinction coefficient to $k < 1 \ 10^{-4}$ around $\lambda = 365\ nm.$ The HSQ thickness is about 
$80 \ nm$ after thermal cure.
Each grating spans over a $100 \times 100 \ \mu m^2$ area,  with the grating grooves oriented parallel to the direction $<1 \bar{1} 00>.$ 

Such a grating couples a wave propagating with a propagation constant $\beta$ parallel to the direction $<11 \bar{2} 0>$ and an energy $E = \hbar \omega$ to an outcoupled light wave at an angle $\theta$ according to the 1st-order diffraction condition:

\begin{equation}
\label{eq:diffraction}
\beta = \frac{\omega}{c} \sin (\theta) \pm \frac{2 \pi}{\Lambda}
\end{equation}

The grating period $\Lambda$ has been chosen for an optimal out-coupling perpendicular to the sample surface in the zero detuning regime (for the $TE0$ mode). A series of gratings with varying periods from $140 \ nm$ to $170 \ nm$ by step of $6 \ nm $ were fabricated, all of them with a measured fill factor $a / \Lambda = 0.43 \pm 0.01,$ where $a$ is the width of each grating stripe.

The TE0 (resp.~ TM0) modal confinement factor is estimated to be of $80 \%$ (resp $65 \%)$ in the GaN guiding core while the overlapping with the grating is typically  lower than $3 \%,$ leading to a relatively weak extraction loss.
The combination of a surface roughness in the order of $1 \ nm$ and a surface morphology made up of steps-and-terraces mounds with a diameter in the order of 1 micrometer or less, results in propagation losses due to surface scattering \cite{Payne_theoretical_1994} estimated to $8 \ cm^{-1}$ at $\lambda = 370 \ nm.$ Thus, this source of losses should only dominate the absorption losses at strongly negative detuning.

\section{\label{sec:exp_dispersion}Experimental polariton dispersions}

The polariton dispersions are measured through angle-resolved micro-photoluminescence under pulsed non-resonant excitation $(400 \ ps$ pulse width, $4 \ kHz$ repetition rate, at $266 \ nm;$  $\approx 30 \ \mu m$ defocused excitation spot). The pulsed excitation allows for an efficient relaxation of the exciton reservoir towards the polariton branch.
The excitation density (typically $1\ mJ.cm^{-2}$ per pulse) is comparable to the threshold for previously investigated polariton lasers based on GaN microcavities \cite{levrat_condensation_2010,zuniga-perez_patterned_2014,jamadi_polariton_2016,jamadi_competition_2019}, even though lasing is not observed here due to the absence of any cavity for the guided polaritons. It is also comparable to the threshold of the recently demonstrated ZnO waveguide polariton laser \cite{jamadi_edge-emitting_2018}.

\begin{figure}
\includegraphics[width=14cm]{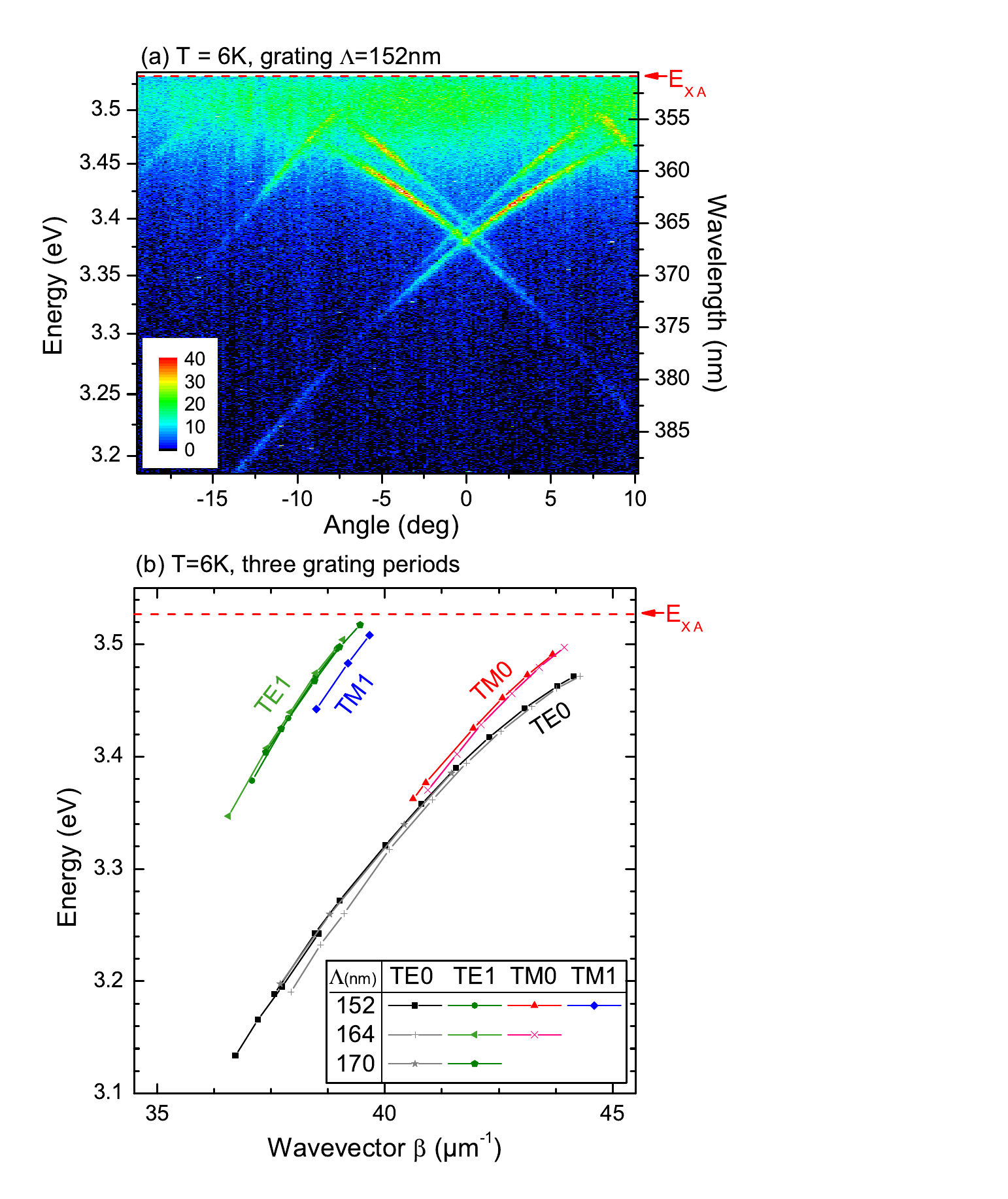}
  \caption{ (a) Angle-resolved photoluminescence $(T=6 \ K)$ of the GaN waveguide diffracted at the grating (period $\Lambda=152 \  nm)$; (b) Experimental exciton-polaritons dispersions of the lower polariton branch extracted from three diffraction gratings, which periods are indicated in the table.}
  \label{fig:dispersion6K}
\end{figure}

The photoluminescence is collected in a Fourier imaging scheme: the back-focal plane of the microscope objective (OFR LMU-20x, numerical aperture 0.40) is imaged onto the entrance slit of the imaging spectrometer and spectrally dispersed ($0.1\ nm$ resolution).
The angle-resolved photoluminescence measured at $T=6 \ K$ for a grating period $\Lambda = 152 \ nm$ is presented in the figure~\ref{fig:dispersion6K}.a. We observe two sets of dispersion curves for positive and negative values of the wavevector $\beta$ along the $z$ propagation axis, i.e. left and right-propagating polaritons: $\beta=E \sin(\theta)  / h c \pm 2 \pi/\Lambda  $ with positive and negative dispersion slopes $E(\theta).$ The broad emission line around $3.5 \ eV$ does not present any structure in Fourier space; it is attributed to localized and donor-bound excitons.

In order to cover a broader range of wavevectors, we present in figure~\ref{fig:dispersion6K}.b the dispersions measured as a function of $\beta$ for grating periods ranging from $152$ to $170 \ nm.$
The identification of the TE and TM modes is deduced from polarization-resolved measurements.
Let us underline that the upper polariton branch, above the exciton energy, is not observed in the experiments due to its strong damping by the absorption of the electron-hole band-to-band transitions, as discussed earlier for GaN and ZnO microcavities \cite{faure_comparison_2008,medard_experimental_2009}. The observed bending of the TE0 and TM0 dispersions when approaching the exciton energy is expected to be the signature of the strong coupling regime between the guided photon and the exciton modes, as will be confirmed later in the discussion. The next section is devoted to the quantitative analysis of these dispersions. Note that the TE1 and TM1, which are mostly confined in the cladding layers, show a much smaller bending.

\section{\label{sec:model}Modeling the polariton dispersion}

{ As discussed in a seminal work on polaritons in microcavities \cite{savona_quantum_1995}, the coupled oscillator model provides a simple implementation of the quantum theory  of the exciton-photon interaction, leading to two coupled modes separated by the Rabi splitting. On the other side, the semi-classical theory is based on a calculation of the optical response through the resolution of the Maxwell equations with a nonlocal linear susceptibility of the active medium. Both approaches provide a complementary understanding of the strong coupling between excitons and photons. In the case of microcavities, the predicted splitting between polariton resonances depends on the probed optical response (reflectivity, transmission, photoluminescence) and it is comparable but not equal to the Rabi splitting deduced from the quantum theory\cite{savona_quantum_1995}.

In the iconic case of a single GaAs quantum well in a microcavity, the simple ``Lorentzian excitonic'' susceptibility provides a relevant description of the active layer, even though all non-resonant contributions are approximated in a frequency-independent background dielectric constant \cite{savona_quantum_1995}. This ``Lorentzian excitonic'' model is implemented in most transfer-matrix simulations of organic and inorganic semiconductor microcavities.

In this section the analysis of the measured polariton dispersions is investigated through four different approaches. The two first models are commonly used in the polaritonics litterature to establish the strong coupling regime in vertical microcavities: the coupled oscillator model (model~A) on one side, and the semi-classical approach with a ``Lorentzian excitonic'' susceptibility (model~B) on the other side. The discrepancy between the measured dispersions and these two  standard models A and~B lead us to consider two more elaborate dielectric susceptibilities for the GaN active layer: the so-called ``Elliott-Tanguy'' model (C) includes all contributions from bound and unbound excitonic states in the dielectric susceptibility; meanwhile, the ``empirical'' model (D) implements the respective contributions from bound excitonic transitions and from all other transitions in the susceptibility as deduced from spectroscopic ellipsometry measurements performed at room temperature \cite{antoine-vincent_determination_2003, mallet_influence_2014}.}
{The fitting of the polariton dispersions with each of the four models will be performed from T=6K (Section~\ref{sec:model}) to T=300K (Section~\ref{sec:temperature}), while the accuracy of the temperature-dependent susceptibility of each model will be tested by comparing it to the experimental ellipsometry measurements performed at T=300K.}
Moreover, the TE and TM modes calculated within isotropic and anisotropic susceptibilities will be further compared.
Finally, we will contrast the four interpretive frameworks and will conclude in the reliability and accuracy of each of them.
 
\subsection{\label{sec:coupled_oscillators} Coupled oscillator model (Model A)}

\begin{figure}
\includegraphics[width=10cm]{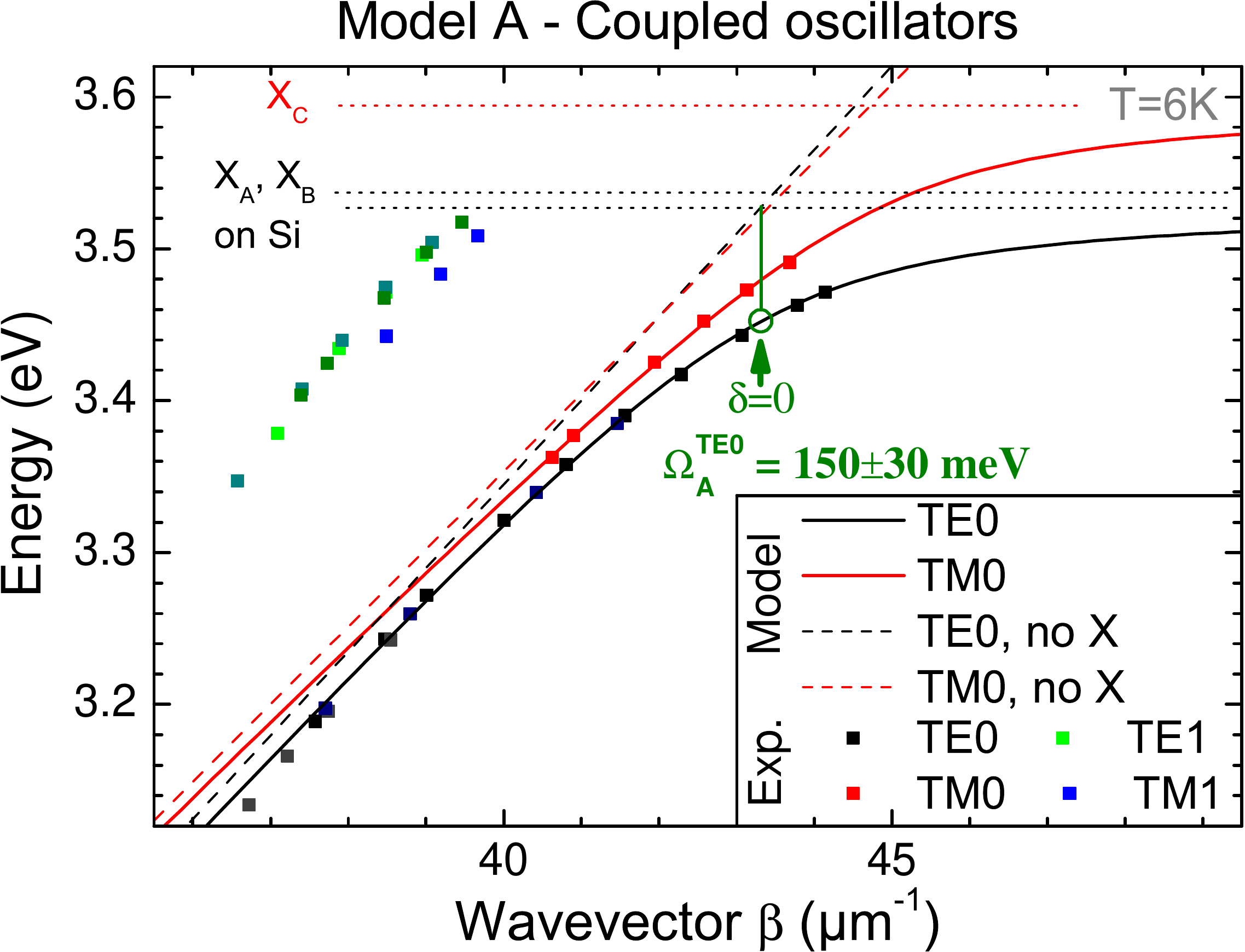}
 \caption{ Model A: Coupled oscillator modeling (plain lines) of the TE0 (black) and TM0 (red) of the polaritons dispersion, the experimental dispersions are shown as square dots. The corresponding bare waveguide modes are indicated as dashed lines, and the exciton energies as horizontal dotted lines. The green open circle indicates the zero-detuning condition, and the green vertical segment represents the half of the exciton-photon coupling strength}
 \label{fig:dispersion_vs_Osc_couples}
\end{figure}

Let us first compare the measured dispersions of the TE0 and TM0 modes with the coupled oscillator model. { As discussed in the Appendix~I, the two optical polarizations can be treated independently. A and B excitons couple to the bare TE0 mode; since they are closely lying, with a splitting much smaller than the expected Rabi splitting, we consider them as a single excitonic oscillator. The C~exciton couples to the bare TM0 mode. For simplicity we don't consider here the coupling of the excitons to higher order optical modes (TE1, TM1 and beyond). For each polarization the coupling is therefore modelled by a wavevector-dependent 2x2 Hamiltonian. Such an approach can be generalized to a broader set of excitonic and photonic modes~\cite{richter_maxwell_2015}.

We assume a linear dispersion of the bare photon mode within the spectral range of interest. The exciton energies are fixed from reflectivity measurements (see Appendix~I). Both calculated TE0 and TM0 polariton dispersions are shown} in Figure~\ref{fig:dispersion_vs_Osc_couples}. The fit over three parameters (the two coefficients of the photon line and the exciton-photon coupling strength) provides an estimate of the exciton-photon coupling strength for the TE0 polariton mode: $\Omega_A^{TE0} = 150 \pm 30 \ meV.$ The large uncertainty on $\Omega_A$ results from the large uncertainty on the group velocity of the bare photon mode, thereby influencing the wavevector $\beta_{\delta0}^A = 43.3 \pm 1 \ \mu m^{-1}$ that corresponds to zero detuning. The same fitting procedure for the TM0~polariton mode leads to $\Omega_A^{TM0} = 150 \pm 50 \ meV,$ the uncertainty being even larger than for the TE0~mode because we do not access large negative detunings in the angle-resolved PL experiments.

\subsection{Waveguide modeling with a susceptibility based on Lorentzian excitonic resonances (Model B)}

The standard approach when investigating the strong coupling in a vertical microcavity consists in comparing the previous coupled oscillator model to transfer matrix simulations \cite{savona_quantum_1995}, in which the photonic modes of the microcavity and the dielectric susceptibility of the exciton active layer (Ref.~\citenum{reveret_modelling_2010} for GaN microcavities) are both taken into account. The equivalent for the waveguide geometry is the resolution of the guided modes in a dielectric slab. Here we choose the resolution for the scalar description of the electromagnetic field either in the TE or in the TM polarization, therefore neglecting any TE-TM coupling. The modes are calculated with CamFR\textregistered, an open-source code implementing the vectorial eigenmode expansion method \cite{bienstman_optical_2001}, under the approximation of decoupled TE and TM eigenmodes { (see Appendix~I)}. This is fully valid for the TE modes in the absence of coupling between A, B and C excitons. Meanwhile this is an approximation for the TM modes, since the electric field of the TM modes may have a small component along the sample plane, therefore coupling to the A and B excitons. This is also an approximation for the TE mode if we take into account the mixing of oscillator strengths of the A, B and C excitons, but this mixing is of the order of $1 \%$ as discussed in the Appendix~I. We can therefore consider that the scalar resolution of the waveguide eigenmodes is a relevant approximation.

\begin{figure}
\includegraphics[width=10cm]{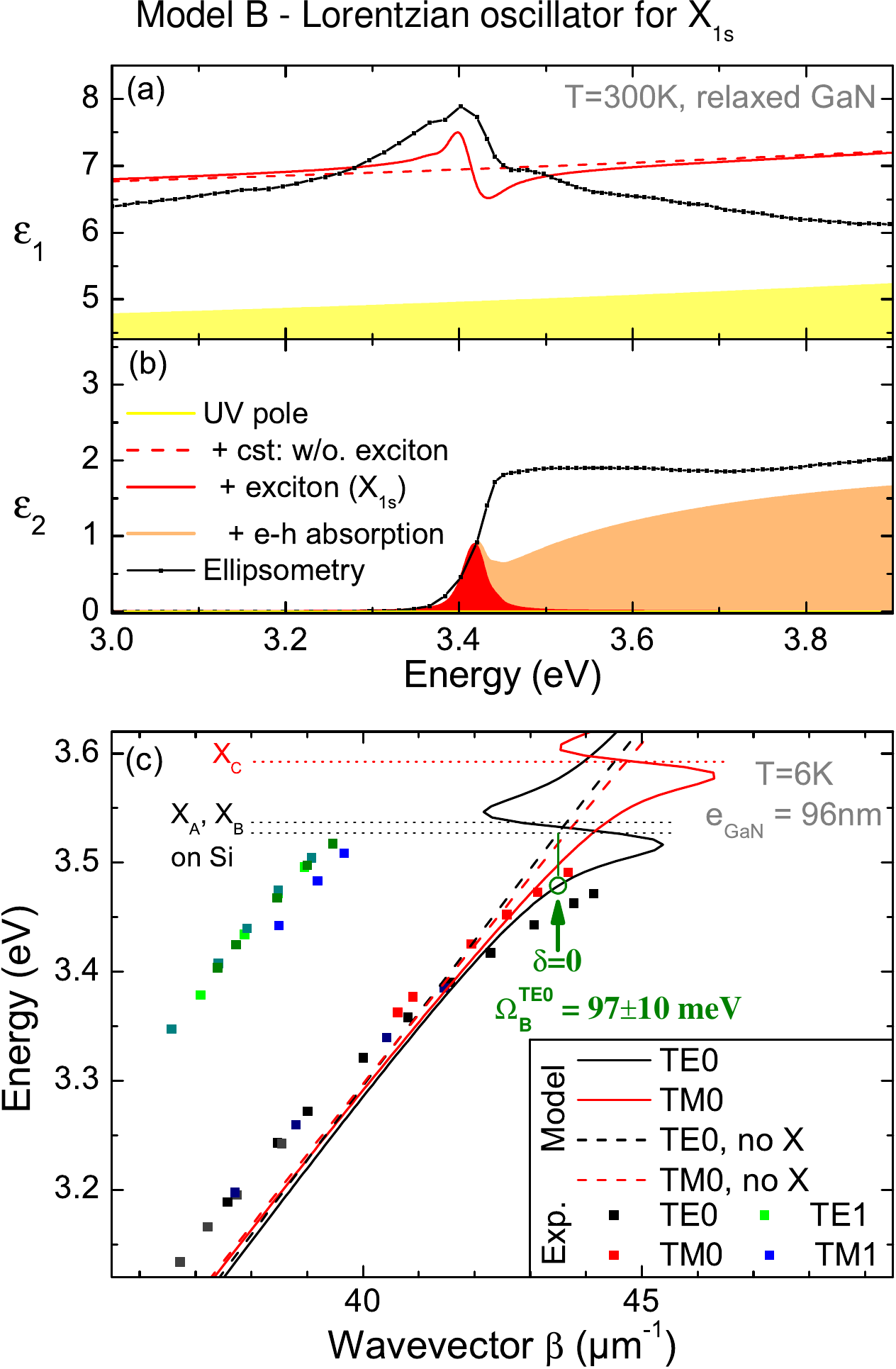}
 \caption{ { Model B based on the nonlocal susceptibility of a single Lorentzian excitonic resonance. (a) Real part and (b) imaginary part of the GaN dielectric susceptibility (in-plane light polarization, TE mode, $T=300 \ K)$: ellipsometry measurements in black line with dots; (red) single Lorentzian oscillator accounting for both A and B~excitonic transitions (including a deep UV Sellmeier pole and a constant shift); (orange) band-to-band contribution to the absorption. (c) Calculated polariton dispersion (plain lines) of the TE0 (black) and TM0 (red).} The experimental dispersions are shown as square dots. The corresponding bare waveguide modes are indicated as dashed lines, and the exciton energies as horizontal dotted lines.}
 \label{fig:dispersion_vs_LorentzX1s}
\end{figure}

Following this approach with the effective background dielectric constant as a free parameter, and considering a typical oscillator strength $40\ 000 \ meV^2$ for each A and B excitons and $80\ 000 \ meV^2$ for the C exciton \cite{julier_determination_1998,butte_room-temperature_2006,reveret_strong_2007,reveret_influence_2008,mallet_influence_2014}, { we compare the calculated and experimental polariton dispersions in Figure~\ref{fig:dispersion_vs_LorentzX1s}.c.} It is seen that the calculated polariton dispersion fails to reproduce the experimental measurement :
(i) The experimental dispersion near the anti-crossing is not reproduced by the model;
(ii) the dispersion at negative detunings is not fitted properly, and the corresponding group velocity of the bare photon mode is $20 \%$ larger than the measured one, even if we account for the dispersive refractive index of the AlGaN and AlN cladding layers. The Rabi splitting extracted from this model is only $97 \pm 10 \ meV$, much smaller than the result of the coupled oscillator model.

{ To unravel the origin of the discrepancy between the experimental and calculated dispersions, we compare for a thick relaxed GaN layer at room temperature the experimental complex dielectric susceptibility $\varepsilon = \varepsilon_1 + i\ \varepsilon_2$ to the ``Lorentzian'' one (Fig.~\ref{fig:dispersion_vs_LorentzX1s}.a,b). The model susceptibility is chosen as the sum of three contributions:
(i) the Sellmeier dispersive susceptibility associated to the deep UV pole of GaN, at about $7 \ eV$ \cite{ shokhovets_determination_2003,kawashima_optical_1997} (yellow curve);
(ii) an empirical rigid shift of $\varepsilon_1;$
(iii) the contribution of the excitonic Lorentzian oscillator to the dielectric susceptibility  (plain red curve). The calculated susceptibility of the GaN material in the absence of the excitonic resonance is the sum of the two first contributions (dashed red line), and is used to calculate the bare guided mode dispersion in the figure~\ref{fig:dispersion_vs_LorentzX1s}.c.

The resulting dielectric susceptibility appears to be a very rough approximation of the experimental one since the obtained complex susceptibility fails to account for three important features: 
(i) the amplitude of the peak in $\varepsilon_1$ near the exciton energy is under-estimated; 
(ii) the strong dispersion of  $\varepsilon_1$ below the bandgap, in the $3.0-3.3 \ eV$ spectral range, is not reproduced; 
(iii) the very strong above-bandgap absorption of GaN, where $\varepsilon_2$ reaches a value of about 2, is absent in the model.
Those three features are strongly related since the step-like increase of $\varepsilon_2$ translates in a strong peak and a long low energy tail for $\varepsilon_1$ through Kramers-Kronig relations.
}

\subsection{Waveguide modeling with the Elliott-Tanguy dielectric susceptibility (Model C)}
\label{sec:Tanguy_susceptibility}
 
\begin{figure}
\includegraphics[width=10cm]{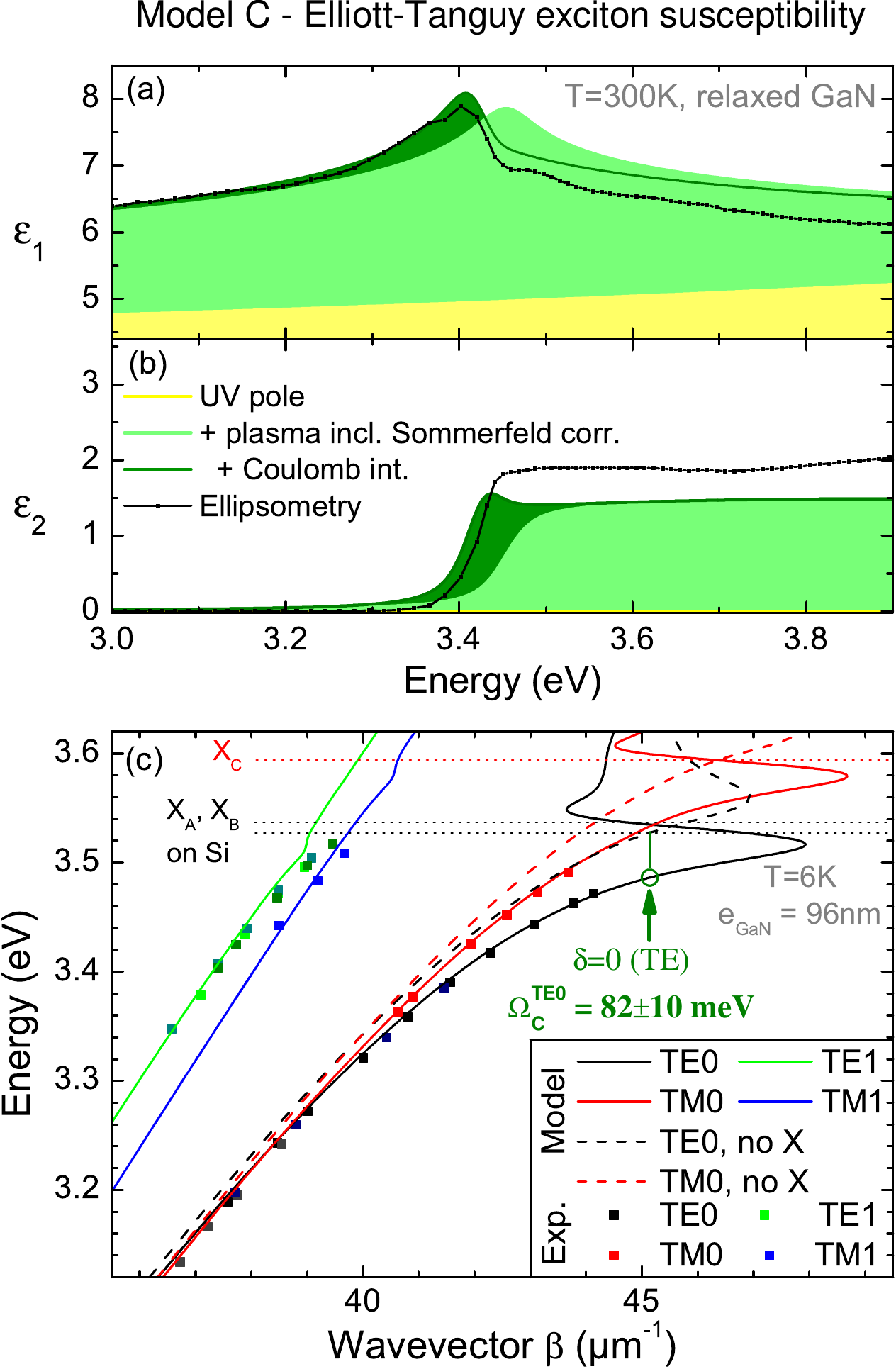}
 \caption{ { Model C based on the Elliott-Tanguy susceptibility. (a) Real part and (b) imaginary part of the GaN dielectric susceptibility (in-plane light polarization, TE mode, $T=300 \ K)$: ellipsometry measurements in black line with dots; Elliott-Tanguy model (including the same deep UV pole as in fig.~\ref{fig:dispersion_vs_LorentzX1s}) with (dark green) and without (light green) Coulomb interaction. (c)} Calculated dispersions (plain lines) of the TE0 (black), TM0 (red), TE1 (green) and TM1 (blue) eigenmodes based on the Elliott-Tanguy model of the dielectric susceptibility, and compared to the experimental dispersions (square dots). The corresponding bare waveguide modes are indicated as dashed lines, and the exciton energies as horizontal dotted lines.}
 \label{fig:dispersion_vs_ElliottTanguy}
\end{figure}

This discrepancy was analytically solved by C.~Tanguy, based on the Elliott model of the absorption of excitons and their diffusion states \cite{elliott_intensity_1957}, in two different cases: the 3D~exciton in bulk semiconductor \cite{tanguy_optical_1995}, as in the present work, and the 2D~exciton in a quantum well \cite{tanguy_analytical_1997}, as in most polaritonic studies. Let us first emphasize that the simple band-to-band absorption, with a $(\sqrt{E-E_G})/E$ dependence in the case of a 3D electron and hole density of states (orange curve for $\varepsilon_2$ in fig.~\ref{fig:dispersion_vs_LorentzX1s}.b), underestimates the experimental absorption near the band-edge~; this evidences the importance of Sommerfeld corrections { related to the electron-hole plasma absorption edge}, as calculated by Elliott \cite{elliott_intensity_1957}. The corresponding $\varepsilon_2$ is presented in fig.~\ref{fig:dispersion_vs_ElliottTanguy}.a,b (dark green curve), with a homogeneous broadening  $(20 \ meV$ at $T= \ 300K)$ and an inhomogeneous broadening $\sigma=10 \ meV$. It reproduces the Heaviside-like absorption-edge, plus the contribution of the excitons as an broadened peak { when Coulomb interaction is taken into account}. Following the work by C.~Tanguy and accounting for the same inhomogeneous broadening, we derive the dispersion of  $\varepsilon_1$. 
No additional constant is required for this approach and a good agreement with the experimental spectroscopic ellipsometry is obtained below bandgap up to the band-edge; a slight discrepancy is observed above bandgap, since $\varepsilon_2$ is still slightly under-estimated.


A major interest of the so-called ``Elliott-Tanguy'' model presented in Fig.~\ref{fig:dispersion_vs_ElliottTanguy}.a,b 
is the possibility to calculate the complex dielectric susceptibility in the presence (dark green curve) and in the absence of Coulomb interaction (light green curve), in the limit of a vanishing Rydberg energy for the exciton as discussed in Ref.~\citenum{tanguy_optical_1995}. This last curve presents a marked peak at the band-edge at $3.45 \ eV$ that is solely related to the band-to-band absorption of GaN (no excitons) \cite{shokhovets_momentum_2005}.

This susceptibility is used to calculate the dispersion of the bare photon modes of the waveguide { within the same CamFR simulation as in the previous section}. The corresponding calculated dispersions for the polariton modes are presented in  figure~\ref{fig:dispersion_vs_ElliottTanguy}. In order to properly fit the experimental dispersions of all TE modes, the oscillator strength of the A, B excitons has been weighted by a factor $0.95,$ and the strength of the UV pole at $7 \ eV$ has been slightly increased by a factor $1.04.$ The GaN thickness is adjusted to $e_{GaN}=96 \ nm,$ very close from the actual thickness $(100 \ nm)$ measured on SEM images. The fit of the polariton dispersions is correct for the modes confined in GaN (TE0) as well as in the AlGaN cladding layer (TE1), thus showing that the refractive index and the waveguide thickness are independently determined. The mode profiles are shown in the Appendix~II.
{ It is noteworthy that, even in the absence of excitons, $\varepsilon_1$  displays a marked peak near the band edge, which leads to a strong bending of the dispersion of the bare waveguide modes.}
The zero-detuning condition $(\delta = 0 \ meV)$ is therefore realized at a wavevector larger than that of the coupled oscillator model, $\beta_{\delta0}^C = 45.1 \pm 0.5 \ \mu m^{-1}$ instead of $\beta_{\delta0} ^A= 43.3 \pm 1 \ \mu m^{-1},$ leading to a smaller value of the exciton-photon coupling $\Omega_C^{TE0} = 82 \pm 10 meV$. The uncertainty is estimated by varying the background refractive index, the exciton oscillator strength and the GaN thickness in the fitting procedure.

\subsection{Empirical dielectric susceptibility from ellipsometry (Model D)}
\label{sec:ellipsometric_susceptibility}

\begin{figure}
\includegraphics[width=10cm]{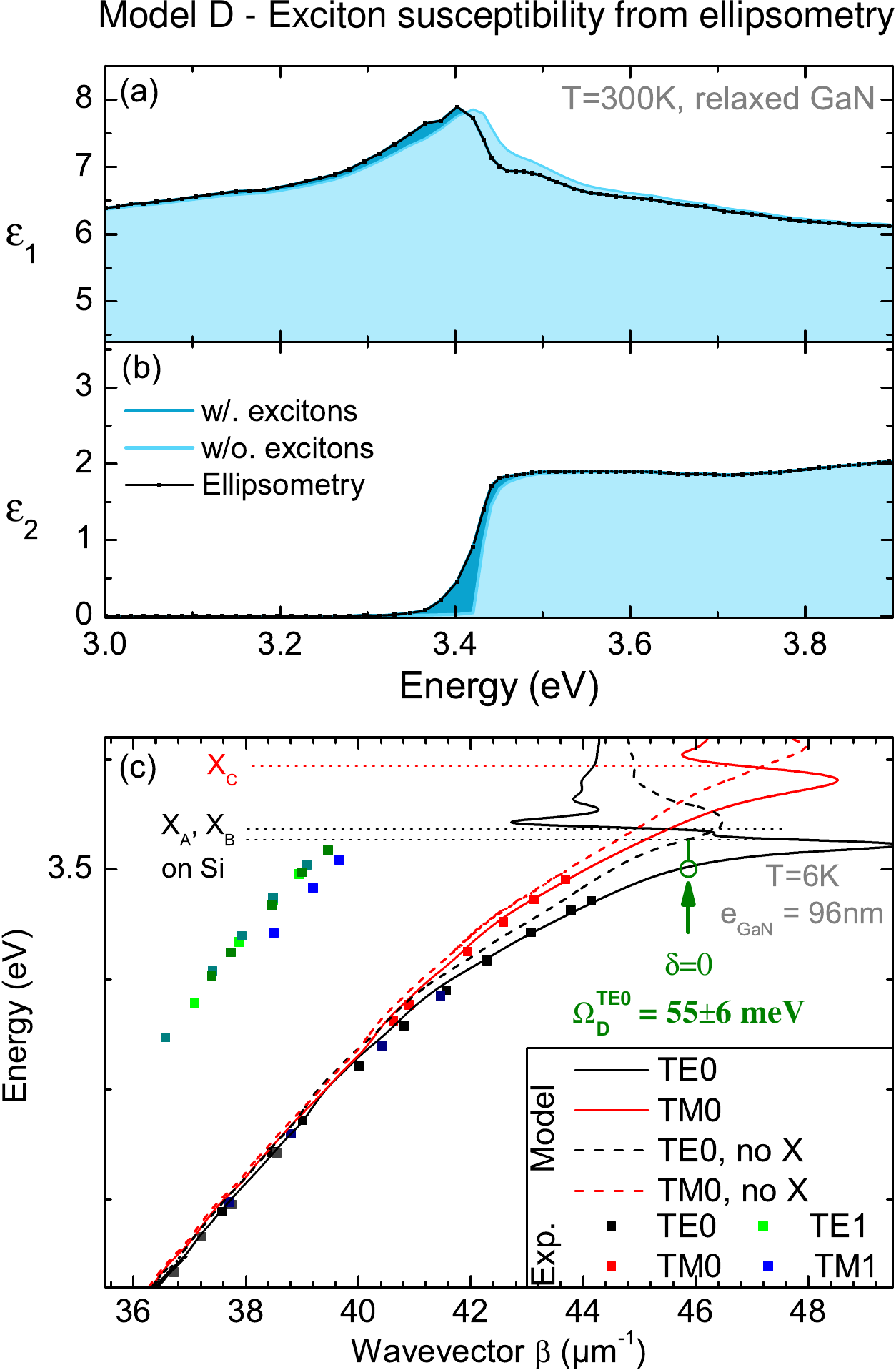}
 \caption{ { Model D based on the empirical susceptibility. (a) Real part and (b) imaginary part of the GaN dielectric susceptibility (in-plane light polarization, TE mode, $T=300 \ K)$: ellipsometry measurements in black line with dots; Susceptibility without excitons (light blue) as the substraction of the experimental measurement and the Lorentzian oscillators. (c) Calculated polaritons dispersion (plain lines) of the TE0 (black) and TM0 (red),} compared to the experimental dispersions (square dots). The corresponding bare waveguide modes are indicated as dashed lines, and the exciton energies as horizontal dotted lines.}
 \label{fig:dispersion_vs_ModelEllipso}
\end{figure}

We have already highlighted that the extraction (or derivation) of a dielectric susceptibility of the GaN layer in the absence of excitons is a critical input for the calculation of the dispersion of the bare photon modes and, as a consequence, {for the deduction of the exciton-photon coupling strength that relates the semi-classical approach to the Rabi splitting obtained in a quantum theory.} We propose here another approach where we assume that the susceptibility measured by spectroscopic ellipsometry (Fig.~\ref{fig:dispersion_vs_ModelEllipso}.a,b, black line, T=300 \ K) is accounting reliably for the excitonic transitions as well as for all other contributions to the background dielectric response of the GaN layer. Starting from the experimental susceptibility, the contribution of excitons to the dielectric constant is mathematically substracted to the experimental data by considering excitons as classical harmonic oscillators. Only A and B excitons in their fundamental (n=1) and excited states (up to n=4) are considered. An overall broadening of $23 \ meV$ is chosen for an optimal substraction of the excitonic contribution to the whole dielectric constant. The energies of A and B excitons are adjusted to fit the absorption front of $\varepsilon_2$ located just below the bandgap energy $(3432 \ meV$ at 300~K). All the excitonic parameters being determined through this procedure, the corresponding contribution of the exciton to $\varepsilon_1$ is substracted to the experimental data. Finally, the complex dielectric constant without the excitonic contributions is deduced initially at room temperature, { and it is then obtained at all temperature by shifting the energies, following the temperature dependence of the bandgap.} The susceptibility at 300~K is presented on Fig.~\ref{fig:dispersion_vs_ModelEllipso}.a,b. 
This approach provides a third estimate of the GaN susceptibility in the absence of excitons.

The corresponding polariton dispersions are presented in Fig.~\ref{fig:dispersion_vs_ModelEllipso}.
The main change compared to the Elliott-Tanguy model concerns the red shift of the band-edge resonance for the TE0 mode in the absence of excitons. This leads to an even larger wavevector for the zero-detuning condition $(\beta_{\delta0}^D = 45.8 \pm 0.5 \ \mu m^{-1}),$ and a much smaller Rabi splitting $\Omega_D^{TE0} = 55 \pm 6 \ meV.$

\subsection{\label{ssec:comparison}Comparison of the four models}

From the above discussion one understands that the choice of the proper model for the dielectric susceptibility of the GaN active layer without excitons has a major impact on the determination of the exciton-photon coupling strength. { The deduction of a single figure of merit of the strong coupling regime ---the Rabi splitting corresponding to the coupled oscillator quantum theory--- is hiding a large part of the complexity of the electronic excitations in the active layer, and of the related dielectric susceptibility.} From a practical point of view, a good quantitative agreement between the experimental and simulated dispersions is not sufficient to secure a reliable estimation of the Rabi splitting. This is due to 
(i) the large oscillator strength of optical transitions in GaN, which leads to a strong  absorption from the excitons as well as from the unbound electron-hole pairs: a constant background value for $\varepsilon_1$ is too rough an approximation; and
(ii) the polariton Rabi splitting being larger than the exciton binding energy, the peak observed in the dielectric susceptibility in the absence of excitons lies in the same energy range as the bare photon mode at the zero-detuning condition. The simultaneous fulfilment of these two conditions is not specific to bulk GaN polariton devices and should also apply to ZnO, InGaN/GaN quantum wells, { and to some perovskites, transition metal dichalcogenides and organic materials} commonly used as active layers in room-temperature polaritonic devices.

Within the four proposed calculation schemes, the coupled oscillator model (A) is clearly over-estimating the coupling strength; model (B) with an { almost} constant group velocity for the bare photon mode and a single $X_{1s}$ Lorentzian oscillator is not able to fit the whole dispersion and slightly over-estimates the coupling strength. The empirical susceptibility based on ellipsometry (model D) provides the most conservative estimate of the Rabi splitting $(55 \pm 6 \ meV);$ finally, the Elliott-Tanguy model (C) proposes a fully analytical expression for the susceptibility and leads to a intermediate value of $82 \pm 10 \ meV.$ { We shall discuss here the discrepancy between models B, C and D, which provide incompatible values of the coupling strength: it is known from Ref.~\citenum{savona_quantum_1995} that for a given set of exciton and photon energies, the splitting between transmission dips in a microcavity strongly depends on the value of the damping parameters for the photon mode (i.e. the mirror reflectivity) and the exciton mode (i.e. the exciton homogeneous broadening).} { Here these two parameters are part of the modelled dielectric susceptibilities through the absorption of the continuum (controlling the lifetime of the bare photon) and the broadening of the excitonic oscillators. The damping of the bare photon mode is strongly varying among the models B, C and D.} { Our attempt to perform the reverse engineering of the dielectric susceptibility in the absence of excitons is based on two very different sets of assumptions in all models~B-D, leading to very different values of the wavevector $\beta_{\delta0}$ at the zero-detuning condition, and of the splitting between the bare photon mode and the polariton mode at the zero-detuning condition. The discrepancy can therefore not be resolved.} While all models do clearly assess the realization of the strong coupling regime in the investigated GaN bulk waveguides, the actual value of the Rabi splitting is strongly model-dependent{; as a single physical quantity, it cannot hold for all the features of the coupling of the optical modes to the excitonic active layer in a regime where the Rabi splitting is larger than the excitonic binding energy, and where the exciton oscillator strength cannot be considered as a perturbation of the dielectric response of the active layer { (in particular for $\varepsilon_2).$}}

\section{\label{sec:TM}General considerations on the anisotropic optical character of GaN active region: TE and TM polariton modes}

The previous calculations were performed as if the GaN bulk material were isotropic. As detailled in Appendix~I, in the case of a GaN waveguide grown along the $c$ axis and taking into account the estimated strain of the GaN-on-Si layer, pure selection rules (i.e. without mixing) are almost perfectly fulfilled: within $1 \%$ accuracy A and B excitons interact with the TE mode only and the C exciton interacts with the TM mode only. The calculation of the waveguide modes in TE and TM polarizations can be thus decoupled \cite{rosencher_optolectronics_2004}. The TE mode only depends on the { in-plane diagonal element of the dielectric tensor $\varepsilon_{\parallel}$ and the calculation is strictly the same as in the isotropic case. The propagation equation related to the TM mode depends on the out-of-plane diagonal element of the dielectric tensor $\varepsilon_{\perp}$ and on the ratio  $\varepsilon_{\parallel}/\varepsilon_{\perp}.$ Indeed, the continuity of the magnetic fields at the interfaces involves also $\varepsilon_{\parallel}$. 
Without considering the excitonic resonances which are connected with optical selection rules, the ratio $\varepsilon_{\parallel}/\varepsilon_{\perp}$ is chosen in accordance with Ref.~\citenum{ shokhovets_determination_2003}.}

The deduced Rabi splitting for the TM0 mode based on the Elliott-Tanguy susceptibility (model~C) and the empirical dielectric susceptibility (model~D) worth respectively $79 \ meV$ and $56 \ meV.$ It is noteworthy that they are almost equal to the ones of the TE0 mode, which is consistent with the fact that the sum of A and B exciton oscillator strengths equals that of the C exciton.

\section{\label{sec:temperature}Temperature dependence of the Rabi coupling}

\begin{figure}
\includegraphics[width=12cm]{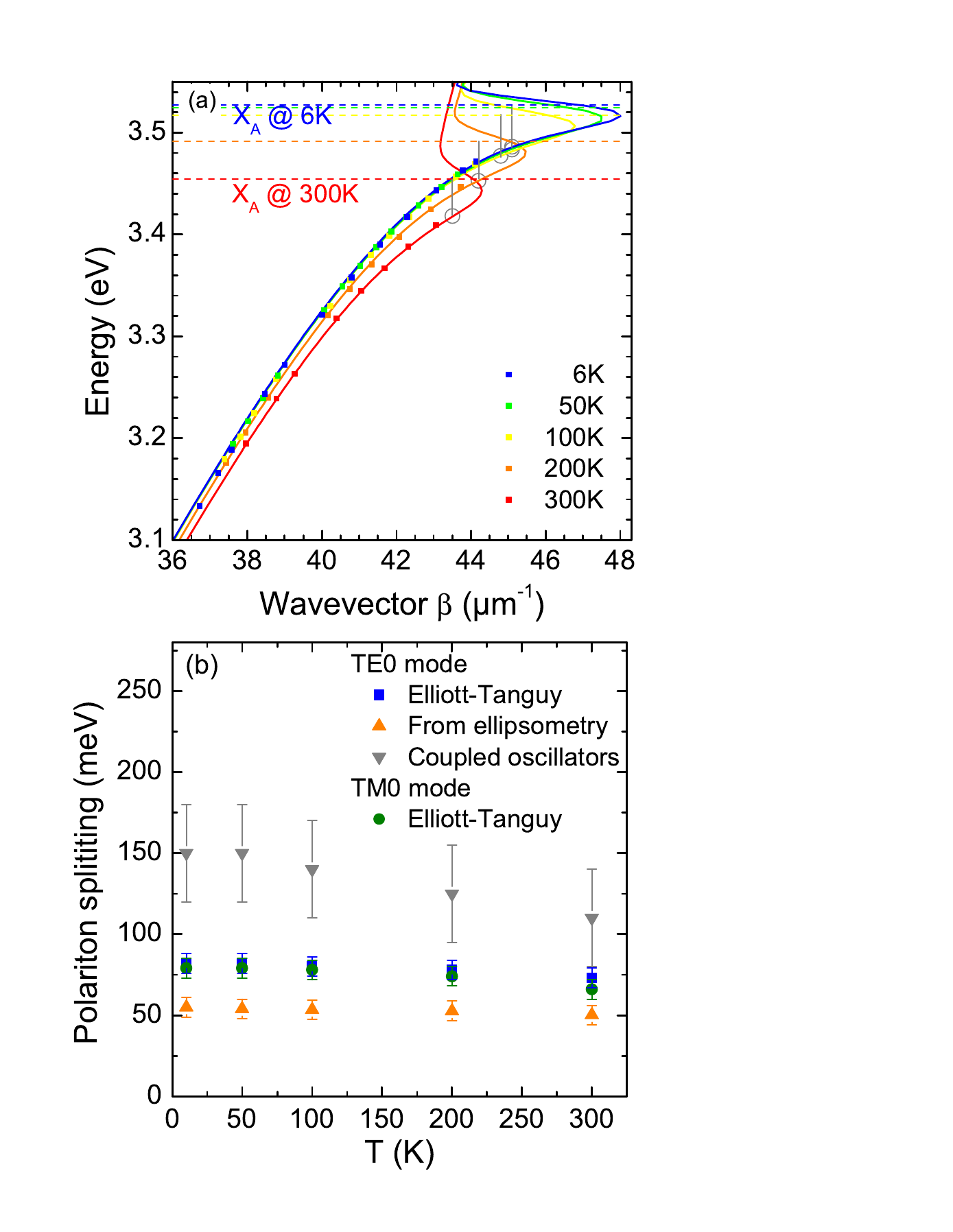}
 \caption{ (a) Experimental dispersions (plain dots) of the TE0 eigenmodes as a function of the sample temperature. The fit by the Elliott-Tanguy model is shown as plain lines with the same colors. The energies of the $X_A$ exciton are presented with dashed horizontal lines; the gray open circles indicate the zero-detuning condition for each temperature, and the gray vertical segments represent the exciton-photon coupling strength $\Omega_{R}/2.$ (b) The corresponding values of  $\Omega_{R}$ are presented for the TE0 and TM0 modes (blue and green symbols), as well as the fitted value for the coupled oscillator model (gray) and the empirical dielectric susceptibility (orange).}
 \label{fig:dispersion_vs_Temperature}
\end{figure}

The experimental polariton dispersions have been measured as a function of the sample temperature from low up to $T= 300 \ K,$ as shown as plain dots in figure~\ref{fig:dispersion_vs_Temperature}.a. All dispersions are very similar, even though the exciton energies (horizontal dashed lines) are shifting to the red as the temperature is increased { \cite{brunner_optical_1997}}. The fit by the Elliott-Tanguy model is presented as plain lines: the homogeneous linewidth is increased from $5 \ meV$ to $20 \ meV$ from $T=6 \ K$ to $300 \ K,$ { \cite{reveret_modelling_2010, trichet_one-dimensional_2011}} while the weight of the exciton contribution to the susceptibility is decreased from $0.95$ to $0.75$ to properly reproduce the measured dispersions; in order to properly fit the photonic part of the dispersions at low wavevectors, it is also required to modify the weight of the Sellmeier deep UV pole by about $10 \%$ from low to room temperature. All other fitting parameters are kept constant across this set of temperature-dependent modeling.

The corresponding values of the Rabi splitting for the waveguide polaritons are presented in Fig~\ref{fig:dispersion_vs_Temperature}.b. We  include the values deduced from the coupled oscillator model, the Elliott-Tanguy model and the empirical model. The Rabi splitting decreases by about $11 \%$ for the TE0 mode, from $82$ to $73 \ meV,$ for the Elliott-Tanguy case, and by $16 \%$ for the TM0 mode, assessing that the strong coupling regime is maintained up to room temperature. Within the ellipsometry-based model~D, the decrease of the Rabi splitting is also small (typically $ 5 \%.)$ Overall, all the models assess the robustness of the strong coupling regime up to room temperature.

\section*{Conclusion}

The dispersion of polaritons propagating in a bulk GaN waveguide is measured from 6K to room temperature, and exhibits a strong negative dispersion coefficient that is a signature of the strong coupling regime. 

{ The analysis of the experimental dispersions is performed through four models of the exciton-photon coupling corresponding to the quantum theory (model~A) and the semi-classical theory (models~B-D), assuming various dielectric susceptibilities for the active layer. We introduce two elaborate models leading to estimates of the exciton-photon Rabi-splitting for TE0 modes of $55 \pm 6 \ meV$ in an empirical approach based on ellipsometry measurements, and $82 \pm 10 \ meV$ in the analytical model, so-called ``Elliott-Tanguy'' model. We show that whereas all models can quantitatively fit the experimental dispersions, they result in very different values of the exciton-photon coupling strengths. This complexity is related to two properties of the GaN active layer: the exciton binding energy is smaller than the coupling strength, and the background dielectric constant cannot be considered as frequency-independent, even at large negative detunings, due to the large absorption above the bandgap. It appears that these two features are common to many materials chosen for their robust strong coupling up to room temperature. { A strong bending of the dispersion in such polaritonic waveguides is not a definitive proof of the strong coupling regime.} Interestingly, the proposed Elliott-Tanguy model can be adapted to polaritonic devices embedding such active layers with a large oscillator strength, including ZnO, InGaN/GaN quantum wells, { and some pervoskites, transition metal dichalcogenides and organic materials}, therefore revisiting the determination of the Rabi splitting in those systems.}

\section*{Acknowledgements} 
 
The authors acknowledge fundings from the French National Research Agency (ANR-15-CE30-0020-02, ANR-11-LABX-0014).

{ \section*{\label{ann:exciton energies}Appendix I - Exciton energies and broadenings}

Let us discuss the chosen exciton energy levels of the bulk GaN active layer, that are the same for all the four models. Due to polarization-dependent selection rules, the TE0 mode is coupled to the excitons A and B. The energy of the exciton A is known from a reflectivity experiment performed on a GaN-on-Si sample with a similar thickness: $E_{XA} = 3.527 \ eV.$ If we assume a biaxial strain for the GaN layer, this corresponds to a deformation $\varepsilon = -1.3 \%,$ leading to a B~exciton energy $E_{XB} = 3.537 \ eV$ and a C~exciton energy $E_{XC}=3.594 \ eV$ \cite{alemu_optical_1998,julier_determination_1998}.
It should be mentioned that this deformation induces almost pure polarization selection rules, since the C~exciton transfers only $1.1 \%$ of its mainly $x$ polarized (along the growth axis, i.e.~TM) oscillator strength to the $y$ and $z$ polarization (in-plane, i.e.~TE), whereas the A and B~excitons transfer the corresponding oscillator strength from the $x$ and $y$ polarizations to the $z$ polarization. 
This small deviation from purely polarized A, B, C excitonic transitions is not included in the presented coupled oscillator model, but it will be accounted for in section~\ref{sec:Tanguy_susceptibility}.

The comparison of experimental and model dielectric susceptibilities are based on room-temperature spectroscopic ellipsometry and reflectivity experiments performed on a series of thick GaN epilayers grown on silicon substrates, following the procedure detailed in Refs.~\citenum{antoine-vincent_determination_2003, mallet_influence_2014}.


The dielectric susceptibility presented in Figs.~\ref{fig:dispersion_vs_LorentzX1s}, \ref{fig:dispersion_vs_ElliottTanguy}, \ref{fig:dispersion_vs_ModelEllipso} is measured and modelled at $T=\ 300K$ while the polariton dispersion is measured at $T=6 \ K.$ Thus, to translate our modelling down to low temperature, the temperature-dependence of the GaN refractive index has been taken into account through Varshni law for energy shifts, and a temperature dependent homogeneous linewidth for the excitonic states has been considered $(5 \ meV$ at $T= \ 6K).$ 
}

\section*{\label{ann:mode profile}Appendix II - mode profiles} 

\begin{figure}
\resizebox{9.5cm}{!}{\includegraphics{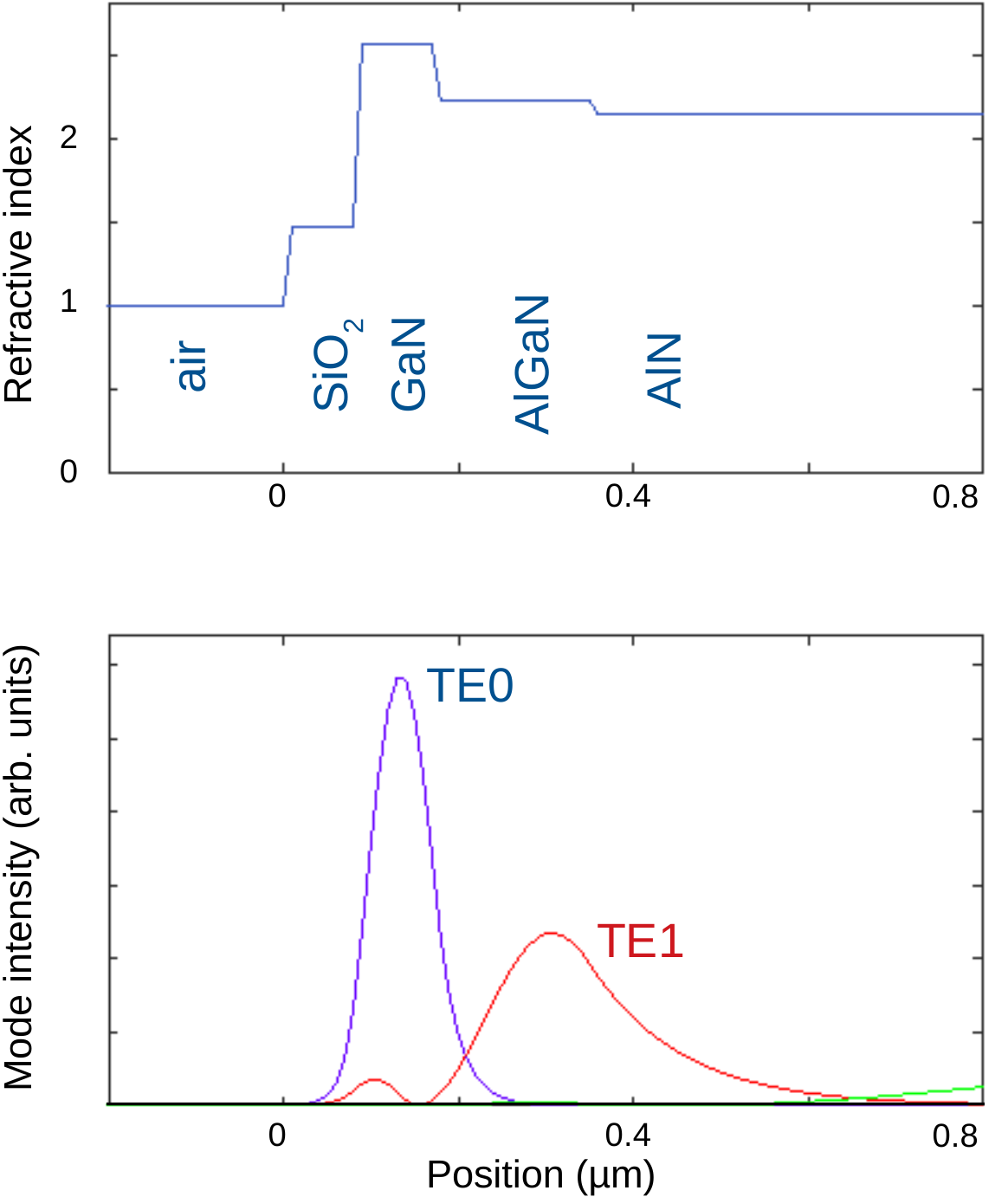}}
 \caption{ Profiles of the refractive index (a) and of the intensity of the two first TE modes (model~C, $T=6 \ K,$ $E=3.487 \ eV).$}
 \label{fig:mode profiles}
 
\end{figure}

The dielectric structure of the GaN waveguide and the calculated eigenmodes are presented on the figure~\ref{fig:mode profiles}, for an optical wave propagating at the energy corresponding to the zero-detuning condition on the figure~\ref{fig:dispersion_vs_ElliottTanguy} (model~C, $T=6 \ K,$ $E=3.487 \ eV).$ The TE0 mode is mostly confined in the GaN active layer $(84 \ \%),$ whereas the TE1 mode is mostly guided in the cladding layers $(4 \ \%$ in GaN, $57 \ \%$ in AlGaN and $35 \ \%$ in AlN). 

\bibliographystyle{./apsrev4-2}
\bibliography{../../Biblio-TGuillet}

\end{document}